\documentclass[amssymb,amsfonts,aps,prb,twocolumn]{revtex4}
\usepackage{graphicx}

\begin{document}

\title{Out-of-plane Enhanced  Magnetic Anisotropy Energy in Ni$_{3}$Bz$_{3}$ molecule}

\date{\today}

\author{T. Alonso-Lanza$^{1,\,}$\footnote{Corresponding author: tomas{\textunderscore}alonso001@ehu.eus}, J. W. Gonz\'alez$^{1,\,}$\footnote{Corresponding author: sgkgosaj@ehu.eus}, F. Aguilera-Granja$^{1,2}$, A. Ayuela$^{1}$}
\affiliation{
$^{(1)}$ Centro de F\'{i}sica de Materiales (CSIC-UPV/EHU)-Material Physics Center (MPC), Donostia International Physics Center (DIPC), Departamento de F\'{i}sica de Materiales, Fac. Qu\'{i}micas UPV/EHU. Paseo Manuel de Lardizabal 5, 20018, San Sebasti\'an-Spain.
\\
$^{(2)}$ Instituto de F\'isica, Universidad Aut\'onoma de San Luis Potos\'i,  78000 San Luis Potos\'i, M\'exico.}

\begin{abstract}
Organometallic complexes formed by transition metals clusters and benzene molecules have already been synthesized, and in selected cases display magnetic properties controlled by external magnetic fields.
We have studied Ni$_n$Bz$_n$ complexes made of nickel atoms surrounded by benzene molecules and here we focus specifically on the magnetic molecule Ni$_{3}$Bz$_{3}$. By means of calculations including relativistic spin-orbit terms, we show that this molecule reveals a large magnetic anisotropy energy of approximately 8 meV, found with the easy axis perpendicular to the metal atoms plane. 
Note that the matching bare Ni$_{3}$ cluster have similar magnetic anisotropy, however the easy axis is in-plane. 
Covering with benzene molecules is thus switching the easy axis from in-plane for Ni$_3$ to out-of-plane for Ni$_{3}$Bz$_{3}$. 
The large out-of-plane magnetic anisotropy of Ni$_{3}$Bz$_{3}$ suggests that this molecule could indeed be used as part in the design of molecular magnetic memories.

{\center{\includegraphics[width=0.5\textwidth]{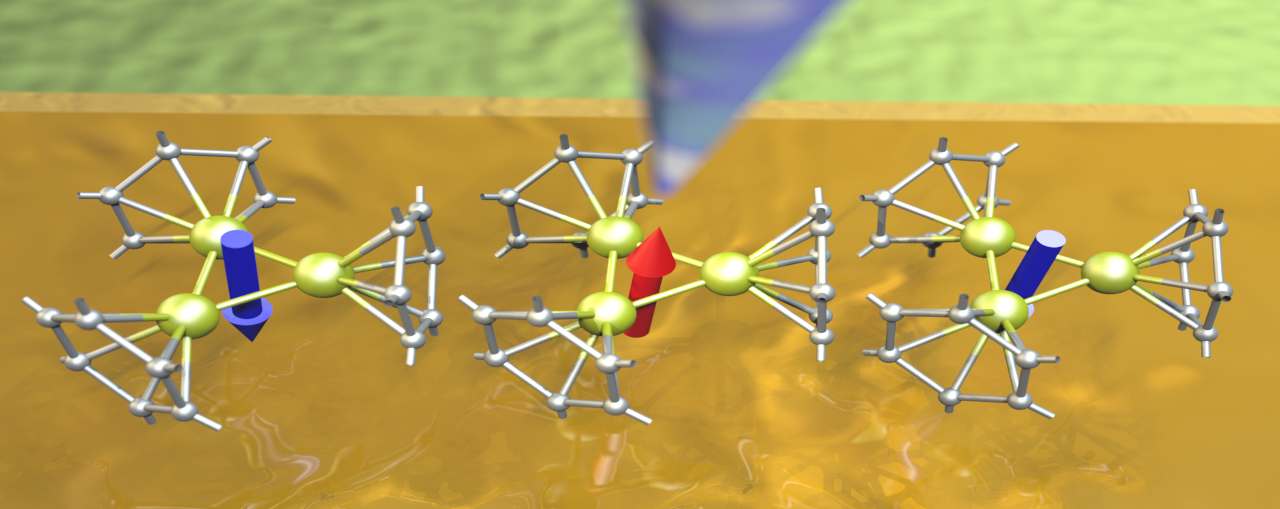}}
\par}
\end{abstract}

\maketitle

\section{Introduction}

The interaction of transition metal atoms with organics molecules have attracted much interest in theoretical and experimental studies since the synthesis of ferrocene \cite{kealy1951new}.   Related complexes known as metallocenes are formed by replacing iron with other transition metals  \cite{pansini2016trends,zeng2014metallocene,yu2016quantum}, and have widespread importance in a wide range of applications, such as spintronics \cite{liu2007cobaltocene,liu2005organometallic,timm2006spin}, biomedical products \cite{ornelas2011application,gomez2012discovery}, and solar cells efficiency \cite{li2014controlled,ishii2014metallocene,park2012hafnium}.
Complexes formed by benzenes (Bz) and first-row transition metal atoms are currently being considered \cite{wedderburn2006geometries,pandey2001electronic,pandey2000unique,muhida2004spin,miyajima2007stern,duncan2008structures}. 
Among those, experiments are discovering and synthesizing a large number of nickel-benzene complexes, which are later characterized by photoelectron spectroscopy \cite{zheng2005photoelectron,kurikawa1999electronic}.
The simplest unit in these compounds, NiBz, places the nickel atom over a benzene hollow site \cite{rabilloud2005geometry,flores2016stability}. The ground state structure for metallocene NiBz$_{2}$ shows the Ni atom attached in a bridge site between carbon-carbon bonds in Bz molecules, that are then being shifted \cite{froudakis2001structural,zhou2006novel,rayane2003electric,flores2016stability}.  However, experiments indicate that large Ni-Bz compounds adopt the so-called riceball structure, in which Ni atoms are wrapped by Bz molecules \cite{kurikawa1999electronic}.

Magnetic effects in benzene transition metal complexes are remarkable and of particular interest for applications. For instance, non-collinear magnetic orders for Co$_3$Bz$_{3}$ have been found \cite{gonzalez2016non}, magnetism depletion is observed in manganese-benzene compounds \cite{alonso2017ultrashort}, and it has been proposed that cobalt dimers on benzene may be magnetic storage bits \cite{xiao2009co}.
Early transition atoms such as Sc, V and Ti when they are placed over benzene molecule enhance magnetic moments, others such as Mn, Fe and Co atoms decrease magnetic moments, and Ni atom magnetic moment is totally quenched \cite{pandey2000unique}. 
Previous works on Ni-Bz complexes \cite{rao2002caging} showed that caging small nickel clusters with benzene molecules fully deplete the magnetism in most of the cases, such as NiBz, NiBz$_{2}$, Ni$_{2}$Bz$_{2}$ and Ni$_{3}$Bz$_{2}$. The magnetism of bare nickel clusters was partially kept for cases, such as for Ni$_{2}$Bz and Ni$_{3}$Bz, 
when nickel-benzene rate is largely above one. As the size of the encapsulated nickel cluster increases the magnetic moment is more easily retained; Ni$_3$Bz$_{3}$ molecule preserves the magnetic moment of the bare nickel cluster. Here we focus on Ni$_3$Bz$_{3}$ because it is the smallest Ni-Bz complex retaining magnetism.

\begin{figure}[thpb]
      \centering 
\includegraphics[width=0.48\textwidth]{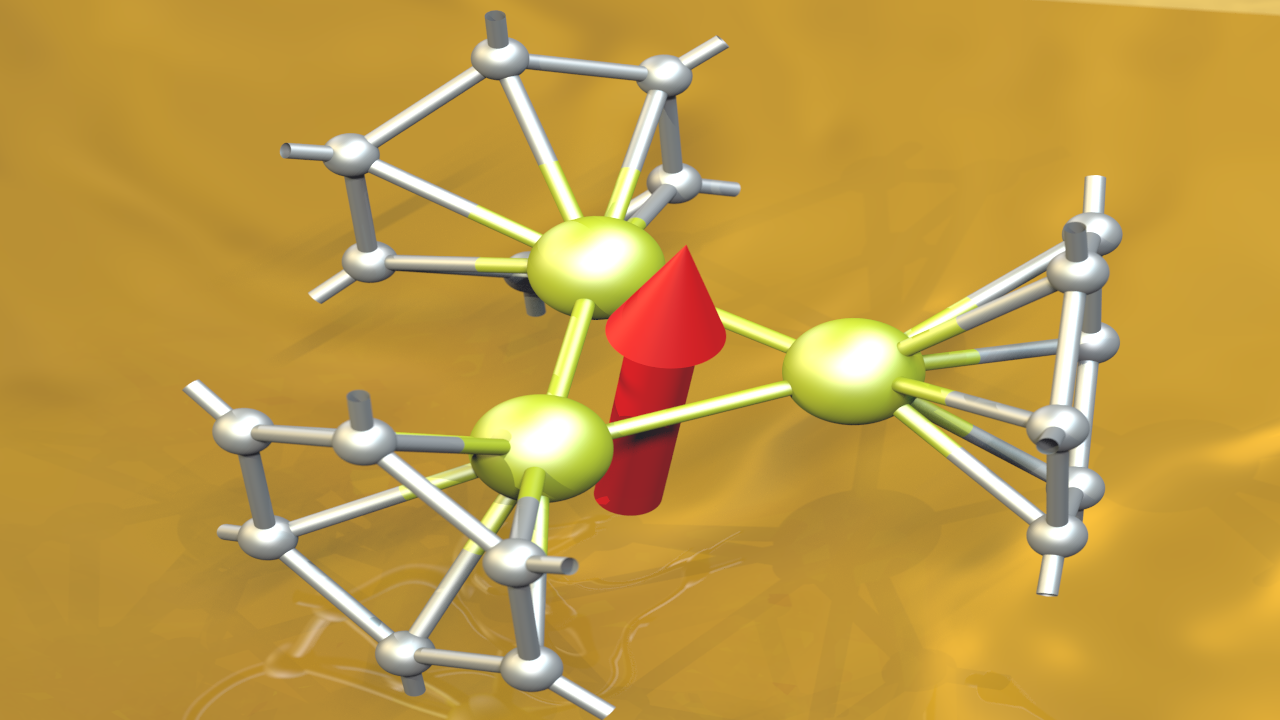}
\caption{\label{figure0} 
Sketch of the ground state geometry of Ni$_{3}$Bz$_{3}$ molecule. The red arrow represents the easy axis of magnetization. The large out-of-plane anisotropy observed in these molecules enables a possible application as molecular magnetic storage units. 
} 
\end{figure}

In this paper we investigate the structural and magnetic properties focusing on Ni$_{3}$Bz$_{3}$ complex by means of density functional calculations including spin-orbit interaction. The complex shows magnetism, keeping the same spin magnetic moment as for Ni$_3$ cluster. First we find the ground state of Ni$_{3}$Bz$_{3}$ molecule among different magnetic arrangements. Benzene molecules strongly affect the magnetic properties of the bare Ni$_3$ counterpart, specifically the magnetic anisotropy.
The two systems show a large magnetic anisotropy energy of approximately $8$ meV, much larger than bulk Ni value in the order of few $\mu$eV per atom. However, the easy axis of magnetization switches from in-plane for Ni$_{3}$ to out-of-plane for Ni$_{3}$Bz$_{3}$ induced by benzene molecules.
The large out-of-plane anisotropy in the Ni$_{3}$Bz$_{3}$ complexes may enable a possible application as molecular magnetic storage units, as sketched in Fig. \ref{figure0}.

\section{Methods}

We perform density functional theory calculations using different methods. 
Spin polarized calculations of Ni-Bz complexes are performed using Vienna ab-initio simulation package  (\textsc{vasp}), based on the projected augmented wave method \cite{blochl1994projector,kresse1999ultrasoft}. For the exchange and correlation potentials we used the Perdew-Burke-Ernzenhof form of the generalized gradient approximation  \cite{perdew1996generalized}.
We relax the structures of Ni$_{3}$Bz$_{3}$ and the Ni$_{3}$ molecules. A cut-off energy of 400 eV for plane wave basis set is used. 
We choose a cubic unit cell with sides 30~\AA{} because such a cell is large to avoid interactions between images. For the calculations we use a $\Gamma$-point sampling. 
The molecule co\-or\-di\-na\-tes are relaxed until the atomic forces are less than 0.006 eV/\AA{}. 
The results were also reproduced using the \textsc{siesta} 
method \cite{soler2002siesta}.
The atomic cores were described by nonlocal norm-conserving relativistic Troullier-Martins pseudopotentials \cite{troullier1991efficient} with non-linear core corrections factorized in the Kleynman-Bylander form \cite{kleinman1982efficacious,alonso2016substitutional}.
We use an electronic temperature of 25 meV and a mesh-cutoff of 250 Ry. 

For the geometry obtained with \textsc{vasp}, we study the effect of the spin-orbit interaction on the different magnetic configurations 
using the all-electron accurate full-potential linearized augmented plane-wave (FP-LAPW) method as implemented in \textsc{elk} software \cite{ELK,singh2006planewaves,sjostedt2000alternative}.
To include the non-collinearity the electronic exchange-correlation potential is treated within local spin density approximation \cite{von1972local}. Wavefunctions, density, and potential in FP-LAPW calculations are expanded in spherical harmonics within spheres and in plane waves in the interstitial region. We use a plane-wave cut-off of $K_{max} R_{Ni}=15$, where $R_{Ni}\approx 1.1$ \AA{} is the nickel muffin-tin radius.
The local magnetic moments are set in particular directions by applying small initial magnetic fields that decrease in each self-consistent loop as to be negligible \cite{singh2006planewaves}.
Calculations use now  $15$ \AA{}  of empty space to avoid interactions between nearest-neighbor cells. 

We are performing state-of-the-art all electron relativistic calculations including non-collinearity, with spin-orbit coupling for valence electrons. Orbital moments are mainly described using the spin-orbit coupling treated self-consistently, which is the source of intrinsic magnetic anisotropy, crystal fields are implicitly included in the calculation. Going beyond the single determinant picture is taken somehow into account using non-collinear calculations for magnetic solids within condensed matter codes, a fact that is considered using mixing determinants within quantum chemistry. It is difficult to say which approach is better, but results using both types might be brought into contact when calculations include the required physics and chemistry. Notice that multideterminant calculations such as CASSCF including all electrons for 3 Ni, 18 C and 18 H atoms seem beyond current computational resources. Last but not least, because we are mainly comparing the ground state properties of two ferromagnetic systems, such as the Ni$_{3}$ and Ni$_{3}$Bz$_{3}$ molecules, density functional theory is perfectly able to deal with the differences \cite{postnikov2006density}.
Specially, when a study of the Ni$_{3}$Bz$_{3}$ molecule at this level of theory has not yet been reported. This study should stimulate experimental work toward the synthesis and characterization of the complex.

\begin{figure*}[thpb]
      \centering
\includegraphics[clip,width=0.9\textwidth,angle=0,clip]{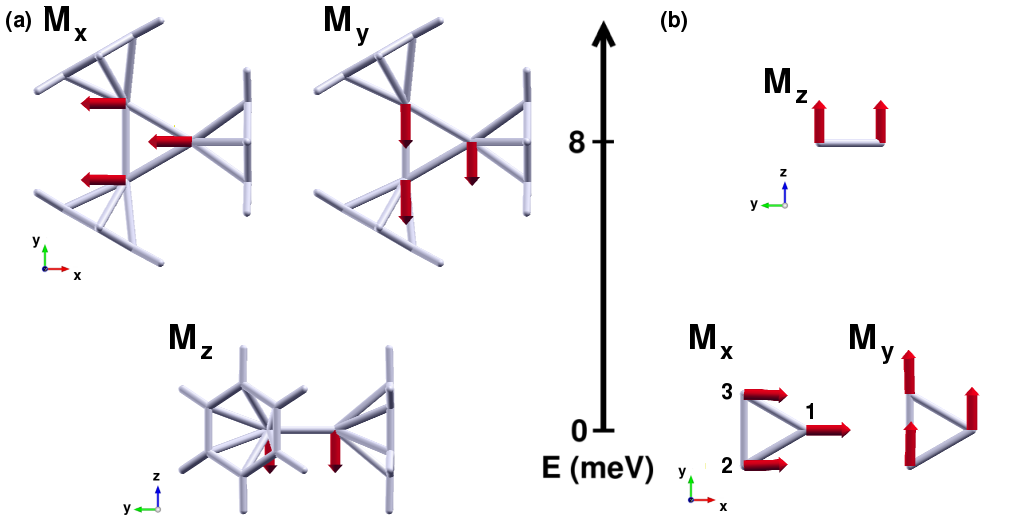}
\caption{\label{figure1} 
Energy of different magnetic configurations for (a) Ni$_{3}$Bz$_{3}$ and (b) Ni$_{3}$. The local Ni spin magnetic moments are denoted by red arrows. Panel of the Ni$_{3}$ M$_x$ configuration includes nickel atoms labeled with numbers, which are used for all the configurations of both Ni$_{3}$ and Ni$_{3}$Bz$_{3}$.} 
\end{figure*}

\section{Results and discussion}

\subsection{Non-spin-orbit calculations}

We fully relax Ni$_{3}$Bz$_{3}$ molecule performing spin-polarized calculations. The ground state is a riceball structure, as shown in Fig. \ref{figure1}. Nickel atoms form an equilateral triangle covered with three benzene molecules, agreing with experiments \cite{kurikawa1999electronic}. 
The nickel-nickel bond lengths are 2.34~\AA{}. 
The Ni-Ni distances for other nickel-benzene complexes are reported to be close, such as 2.40~\AA{} for Ni$_{2}$Bz$_{2}$, 2.45~\AA{} for Ni$_{3}$Bz$_{2}$ \cite{rao2002caging}, and 2.34~\AA{} for Ni$_{2}$Bz$_{2}$ \cite{zhou2006novel}. 
\footnote{The distances between each nickel and the six nearest carbon atoms of the benzene molecule have about 2.21~\AA{}, a value which should be compared well with the 2.04-2.21~\AA{} interval for Ni$_{2}$Bz$_{2}$, and with 2.08~\AA{} for Ni$_{3}$Bz$_{2}$ \cite{rao2002caging} and 2.14-2.34~\AA{} for Ni$_{2}$Bz$_{2}$ \cite{zhou2006novel}, cases in which the nickel atom is not even $\eta^{6}$-bonded. The carbon-carbon bond length is 1.41~\AA{} as found in experiments 1.40~\AA{} \cite{stoicheff1954high}.}
In comparison, the fully relaxed structure for the bare Ni$_{3}$ cluster is an equilateral triangle of side 2.20~\AA{}. The nickel-nickel bonds with Bz molecules enlarge by 0.14~\AA{} (6$\%$). 
Furthermore, the Ni$_{3}$Bz$_{3}$ state is ferromagnetic, found $0.2$ eV below a spin compensated state. The total spin magnetic moment of Ni$_{3}$Bz$_{3}$ has $2$ $\mu_{B}$ with equal contribution from each of the three nickel atoms.  Note that for the bare Ni$_{3}$, a total spin magnetic moment of 2 $\mu_{B}$ is also reproduced, agreeing with previous reports \cite{michelini1998density,papas2005homonuclear,castro1997structure,cisneros1999dft,xie2005first,arvizu2007assessment,michelini2004density}. 
Summing up, Ni in the Ni$_{3}$Bz$_{3}$ metallocene complex is in an oxidation state zero with short Ni-Ni bond lengths and showing ferromagnetic Ni-Ni coupling. The local electronic configuration that in Ni isolated atom is 3d$^8$4s$^2$ becomes close to 3d$^9$4s$^1$ in the complex.

\subsection{Spin-orbit calculations}

\subsubsection{Energy and spin magnetic moments}

We first explore whether collinear or non-collinear arrangements are preferred by computing different solutions of the local Ni spin magnetic moments. 
We start with different non-collinear moments in Ni atoms, such as in-plane radial, 
in-plane tangential, and along \textit{xz}-diagonals. All these configurations converge to collinear local magnetic moments.
The total spin magnetic moment of the ground state of Ni$_{3}$Bz$_{3}$ remains as 2 $\mu_{B}$ with equal contribution from each Ni atom being ferromagnetically aligned, similar to the spin-polarized results discussed above. 

We then include the effects of the relativistic spin-orbit interactions. We align the Ni magnetic moments collinear and sample different in-plane magnetic orientations, for instance along positive and negative directions in the \textit{x}- and \textit{y}-axes. Our results show that solutions having the Ni moments in-plane are nearly degenerate, with energy differences smaller than $0.1$ meV.
We also calculate the case with the Ni magnetic moments along the z-axis perpendicular to the Ni triangle. In fact, this magnetic solution is the ground state, being 8 meV lower in energy than the in-plane solutions. 
The local spin magnetic moments are decomposed around Ni atoms as shown in Fig. \ref{figure1}. It is noteworthy that our results reveal a large magnetic anisotropy energy of $8$ meV with an easy axis out-of-plane, and a hard plane formed by the three Ni atoms. We define the magnetic anisotropy energy as the difference in energy between in-plane and out-of-plane configurations. The magnetic moments of Ni surrounded by Bz molecules suggest that Ni$_{3}$Bz$_{3}$ complexes deposited on metallic surfaces could be used as units in magnetic memories useful in data storage applications \cite{choi2017atomic,chappert2007emergence,rocha2005towards}, although further calculations to evaluate the effect of the substrate are needed.

For comparison we repeat the analysis for the bare Ni$_3$ molecule to show that benzene molecules are key to get the out-of-plane magnetic configuration. The ground state of Ni$_{3}$ is an equilateral triangle with a total spin magnetic moment of 2$\mu_{B}$.
We calculate several arrangements for the local spin magnetic moments by looking for different non-collinear solutions. Nevertheless, our results show that the Ni$_3$ triangle has collinear magnetic moments, as for Ni$_{3}$Bz$_{3}$.
Then, we explore the solutions with in-plane and out-of-plane Ni moments. The ground state of Ni$_3$ is nearly degenerated with several in-plane configurations, with the local Ni magnetic moments pa\-ra\-llel pointing either in the \textit{x}- or \textit{y}-axis. These in-plane states have similar total energies within an interval of 0.2 meV.
The out-of-plane Ni$_3$ magnetic configuration is about few meV higher in energy than the in-plane Ni$_3$ configurations.
This result is reversing the magnetic configuration order found for the Ni$_{3}$Bz$_{3}$ molecule. 
The addition of Bz molecules to Ni$_{3}$ molecule is thus switching the magnetization from the Ni$_3$ easy xy-plane to the Ni$_{3}$Bz$_{3}$ easy z-axis.

\subsubsection{Orbital magnetic moments and experimental considerations}

\begin{figure*}[thpb]
      \centering
\includegraphics[clip,width=0.9\textwidth,angle=0,clip]{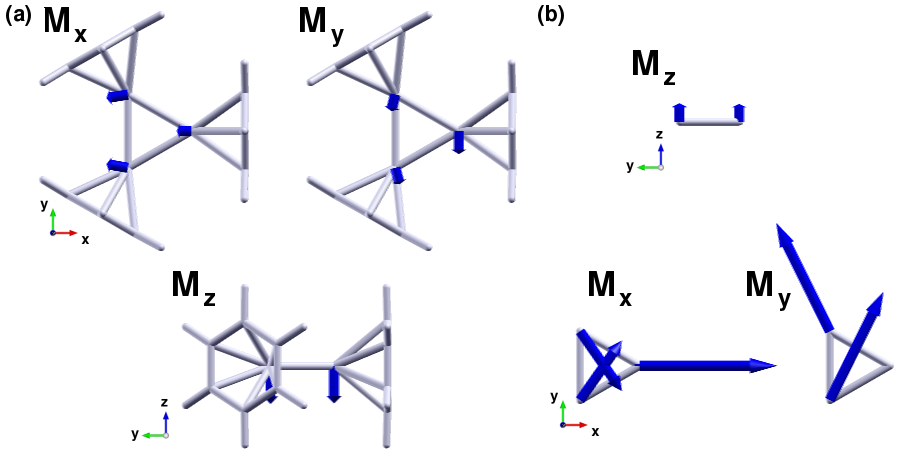}
\caption{\label{figure2} 
Orbital magnetic moments in Ni atoms represented by the blue arrows for (a) Ni$_{3}$Bz$_{3}$ and (b) Ni$_{3}$. A factor $\times 10$ is used when comparing to the local spin magnetic moments of Fig. \ref{figure1}.}
\end{figure*}

We next have to consider that the total magnetic moment can be decomposed in two contributions, namely the spin magnetic moment and the orbital magnetic moment. We have addressed the spin magnetic moments of Ni atoms above, and we focus on the orbital magnetic moments. The values of the $L$, $S$ and $J$ operators in each Ni atom of the Ni$_{3}$ and Ni$_{3}$Bz$_{3}$ systems are given in the Supplemental Material. Figure \ref{figure2} (a) shows the orbital magnetic moments $\vec{L}$ for the different magnetic configurations of the Ni$_{3}$Bz$_{3}$ molecule.  The local Ni orbital magnetic moments nearly follow the direction of the local spin magnetic moments shown in Fig. \ref{figure1}. For the in-plane configurations, the orbital moments of the (2) and (3) Ni atoms are slightly tilted with respect to the local spin magnetic moments.

In comparison, the out-of-plane configuration of Ni$_{3}$ has the orbital magnetic moments still aligned with the spin magnetic moments. However, the in-plane orbital magnetic moments of Ni$_{3}$ are larger than in Ni$_{3}$Bz$_{3}$, and rotate from the direction of Ni local spin magnetic moments. 
For the Ni$_{3}$ in-plane configurations there is even a Ni atom with a zero orbital magnetic moment. The orbital moments are averaged within a sphere centered on each Ni atom, so that regions with different moment direction around a Ni atom could cancel and give a total value of nearly zero Ni moment, as seen in the (1) atom of the M$_y$ configuration in Fig. \ref{figure2}.

The spin and orbital magnetic moments of the isolated Ni cluster cations in gas phases with 7-17 atoms have been analyzed using X-ray magnetic circular dichroism (XMCD) spectroscopy \cite{meyer2015spin}. The orbital magnetic moments for these clusters are within the interval 
0.25--0.43 $\mu_{B}$/atom, much larger than the reported 0.06 $\mu_{B}$/atom for bulk Ni \cite{vogel1994polarization,chen1991exchange}.
In agreement with experiments \cite{meyer2015spin}, we find that the local orbital magnetic moments in Ni$_3$ in-plane configurations have 
values within the interval 0.23--0.44 $\mu_{B}$. The orbital magnetic moments increase for small clusters because the crystal field effect decreases in comparison to bulk. However, when comparing to the nickel atom with $\mu_{L}=3$ $\mu_{B}$, the orbital magnetic moments are considered to be notably quenched, decreasing to 5-25$\%$ of the atomic value \cite{kittel2005introduction,stohr2007magnetism,langenberg2014spin}. For the energetically less stable out-of-plane Ni$_3$ configuration, the $\mu_L$ values decrease even further to 0.06 $\mu_{B}$. 

The out-of-plane Ni$_{3}$Bz$_{3}$ has slightly larger orbital moments of about 0.12 $\mu_{B}$. The in-plane configurations of Ni$_{3}$Bz$_{3}$ also have orbital magnetic moments in the interval 0.05--0.07 $\mu_{B}$, like in bulk Ni \cite{vogel1994polarization,chen1991exchange}. 
The benzene molecule addition makes Ni clusters to behave more like bulk Ni because the local orbital magnetic moments decreases \cite{vogel1994polarization,chen1991exchange}.
Our results agree with previous reported trends \cite{kittel2005introduction,stohr2007magnetism}: for bare small nickel clusters orbital magnetic moments are enhanced, and covering with benzenes quenches the orbital magnetic moments because of the induced crystal field, and align them with the spin magnetic moments.
Furthermore, we find that the most stable configurations, namely out-of-plane for Ni$_{3}$Bz$_{3}$ and in-plane for Ni$_{3}$, are related to the largest values of the orbital magnetic moments. This finding is in agreement with the reported relationship between orbital magnetic moments and magnetic anisotropy energy \cite{bruno1989tight,liz2003nanoscale,hopster2006magnetic}.

\subsection{Electronic structure}

\begin{figure*}[thpb]
      \centering
\includegraphics[clip,width=\textwidth,angle=0,clip]{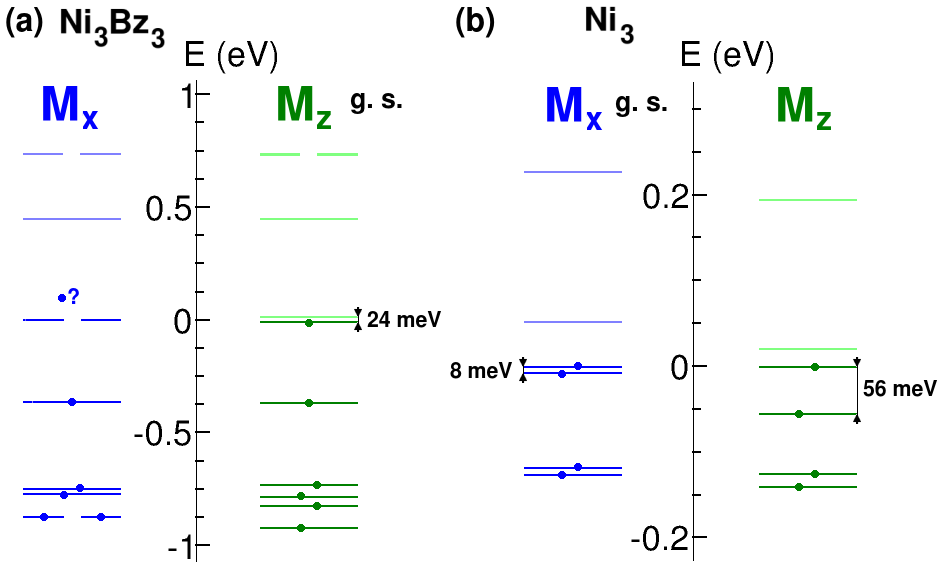}
\caption{\label{figure3} 
Electronic levels near the Fermi energy obtained in spin-orbit (SO) calculations for the (a) Ni$_{3}$Bz$_{3}$ and (b) Ni$_{3}$ molecules.}
\end{figure*}

To understand the effect of the benzene molecules addition, we further study the electronic structure of the Ni$_{3}$Bz$_{3}$ and Ni$_{3}$ clusters.
We comment on the spin-orbit calculations to investigate the magnetic anisotropy change upon magnetic fields; non-spin-orbit calculations are commented in the Supplemental Material.
Figure \ref{figure3}(a) shows the levels of the M$_x$ and M$_z$ configurations for magnetic fields in the \textit{x} and \textit{z} directions, respectively. The M$_y$ configuration behaves similarly to M$_x$.
For the in-plane configurations, either along M$_x$ or M$_y$, there are two degenerated levels near the Fermi energy, sharing the less bonded electron.
Nevertheless, for out-of-plane magnetic moments M$_z$, these two levels near the Fermi level are no longer degenerated. They present a splitting of more than 24 meV, as indicated by arrows in Fig. \ref{figure3}(a), so the electron fills the lower energy level.
This degeneration breaking stabilizes the out-of-plane Ni$_{3}$Bz$_{3}$ configuration (M$_z$) with respect to in-plane cases (M$_x$ or M$_y$) by more than 8 meV.

Next, we compare with the corresponding Ni$_{3}$ levels.
Figure \ref{figure3}(b) shows the levels of the M$_x$ and M$_z$ configurations for magnetic fields in the \textit{x} and \textit{z} directions, respectively. The M$_y$ configuration behaves similarly to M$_x$.
Regarding stability, Ni$_{3}$ in-plane configurations are 8 meV more stable than the out-of-plane configuration, in contrast to Ni$_{3}$Bz$_{3}$.
The in-plane configurations have the two levels near the Fermi energy fully occupied and almost degenerated, with a small splitting of 8 meV. 
In the out-of-plane case the splitting between the two fully occupied levels becomes as large as 56 meV, a value that is about double than for Ni$_{3}$Bz$_{3}$. Out-of-plane magnetic field favors the splitting, as for Ni$_{3}$Bz$_{3}$. 

Nevertheless, there is an important difference between the Ni$_{3}$Bz$_{3}$ and Ni$_{3}$ cases; two electrons fills these two levels for Ni$_{3}$ while only one electron is available for the Ni$_{3}$Bz$_{3}$ couple of levels. 
For Ni$_{3}$Bz$_{3}$ in-plane configurations, the two levels near the Fermi energy share an electron, which is a destabilizing factor. The out-of-plane magnetic field splits these two levels enabling complete filling of one of them, stabilizing the Ni$_{3}$Bz$_{3}$ molecule. This explains the switch of the easy axis between Ni$_{3}$ and Ni$_{3}$Bz$_{3}$.

Finally, we searched for the cause of the difference number of electrons in these two levels near the Fermi energy between Ni$_{3}$ and Ni$_{3}$Bz$_{3}$. 
The role of benzenes reveals as key for the magnetic change. 
The addition of benzene molecules leads to charge transfer towards nickel atoms\footnote{Considering the charge transfer under the Mulliken scheme, we find that for the Ni$_{3}$Bz$_{3}$ molecule, each nickel atoms gains more than 0.2 electrons. In total, the three nickel have 0.65 electrons, in agreement with findings for molecule Co$_3$Bz$_3$  \cite{gonzalez2016non}.} and symmetrizes the system, stabilizing and filling a 4\textit{s} level for Ni$_{3}$Bz$_{3}$, which is unoccupied for Ni$_{3}$. Consequently, the couple of degenerated levels at the Fermi energy for Ni$_{3}$Bz$_{3}$ becomes half-occupied by an electron.

\section{Conclusions}

We here performed spin-orbit calculations for organometallic compounds containing benzene molecules and nickel atoms. We find that molecules Ni$_{3}$Bz$_{3}$ begin to behave as small magnets with all the Ni atom moments being collinear. The effect of benzene molecules is to switch the easy axis of magnetization from in-plane (for the bare nickel cluster Ni$_{3}$) to out-of-plane (for molecule Ni$_{3}$Bz$_{3}$). The in-plane magnetic configurations for molecule Ni$_{3}$Bz$_{3}$, being almost degenerated, are 8 meV higher in energy than the out-of-plane magnetic configuration, a finding that shows a large magnetic anisotropy energy. The out-of-plane easy axis correlates with the large values of the Ni orbital magnetic moments. The switching of the easy axis between the bare and covered Ni$_3$ clusters is explained using the different occupation of two degenerated levels near the Fermi energy, which are split by an applied perpendicular out-of-plane magnetic field.  
Thus, small Ni$_{3}$Bz$_{3}$ magnetic molecules could be of use in applications as magnetic memories because of the large magnetic anisotropy energy with the out-of-plane easy axis, magnetic properties that are further ensured by the protection provided  by the embedding of benzene molecules.

\section*{Acknowledgement}
This work was part financed by the Project FIS2016-76617-P of the Spanish Ministry of Economy and Competitiveness MINECO, the Basque Government under the ELKARTEK project (SUPER), and the University of the Basque Country (Grant No. IT-756-13). TA-L acknowledges a grant provided by the MPC Material Physics Center - San Sebasti\'an. The authors also acknowledge the technical support of the DIPC computer center. We acknowledge Prof. Andrei Postnikov for reading the manuscript.



\pagebreak

\begin{center}
\textbf{\large Supporting Information: Out-of-plane Enhanced  Magnetic Anisotropy Energy in Ni$_{3}$Bz$_{3}$ molecule}
\end{center}
\setcounter{figure}{0} 
\setcounter{section}{0} 
\setcounter{equation}{0}
\setcounter{page}{1}
\renewcommand{\thepage}{S\arabic{page}} 
\renewcommand{\thesection}{S\Roman{section}}   
\renewcommand{\thetable}{S\arabic{table}}  
\renewcommand{\thefigure}{S\arabic{figure}} 
\renewcommand{\theequation}{S\arabic{equation}} 

\section{Magnitudes of the quantum numbers}

Tables \ref{table1} and \ref{table2} include the magnitudes of the angular moments $L$, $S$ and $J$ for Ni$_{3}$ and Ni$_{3}$Bz$_{3}$. It is noteworthy that the $L$ values for Ni$_{3}$  in in-plane configurations are comparable to the $S$ values. The addition of benzene molecules decreases the $L$ value. For Ni$_{3}$Bz$_{3}$ and Ni$_3$, the largest local orbital magnetic moments are found for the most stable out-of-plane and in-plane configurations, respectively.

\begin{table}[h]
\small
\caption{\ Values of the $L$, $S$ and $J$ angular moments in each of the atom spheres for the three different magnetic configurations found for the Ni$_{3}$ molecule.}
\label{table1}
\begin{tabular}{|l|lll|lll|lll|}
\hline
 &  &  M$_x$  &  &   & M$_y$&  &   &  M$_z$&\\
\hline
 & 1  & 2 & 3 &  1  &  2 &  3& 1  & 2 & 3\\
\hline
$L$ & 0.44 & 0.23 & 0.23 & 0.01 & 0.39 & 0.39&  0.06 &  0.06& 0.06 \\
$S$ & 0.34 & 0.34 & 0.34 & 0.33 & 0.34 & 0.34&  0.33 &  0.33& 0.33 \\
$J$ & 0.78 & 0.51 & 0.51 & 0.34 & 0.71 & 0.71&  0.39 &  0.39& 0.39 \\
\hline
  \end{tabular}
\end{table}

\begin{table}[h]
\small
\caption{\ Values of the $L$, $S$ and $J$ angular moments in each of the atom spheres for the three different magnetic configurations found for the Ni$_{3}$Bz$_{3}$ molecule.}
\label{table2}
\begin{tabular}{|l|lll|lll|lll|}
\hline
 &  &  M$_x$  &  &   & M$_y$&  &  &M$_z$  &\\
\hline
 & 1  & 2 & 3 &  1  &  2 &  3& 1  & 2 & 3\\
\hline
$L$ & 0.05 & 0.07 & 0.07 & 0.07 & 0.05 & 0.05& 0.12 & 0.12 & 0.12 \\
$S$ & 0.34 & 0.35 & 0.35 & 0.34 & 0.35 & 0.35& 0.34 & 0.35 & 0.35 \\
$J$ & 0.39 & 0.42 & 0.42 & 0.41 & 0.40 & 0.40& 0.46 & 0.47 & 0.47 \\
\hline
  \end{tabular}
\end{table}

\section{Non-spin-orbit electronic structure}

\begin{figure*}[thpb]
      \centering
\includegraphics[clip,width=0.95\textwidth]{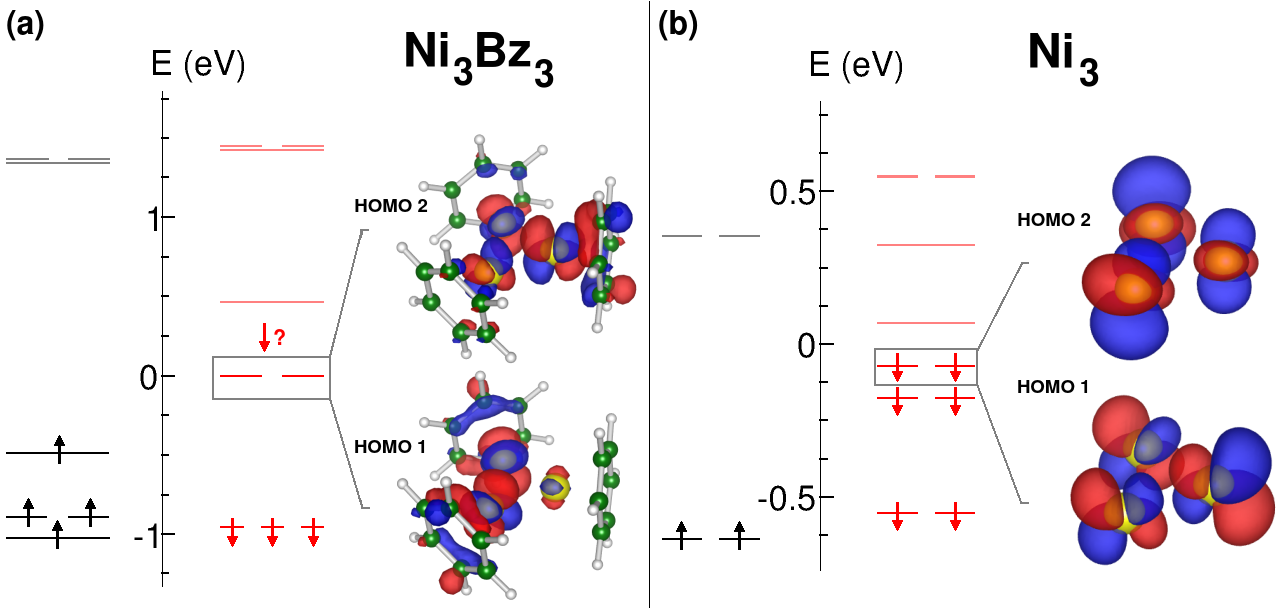}
\caption{\label{figure3b} 
Electronic levels obtained with non spin-orbit calculations for the (a) Ni$_{3}$Bz$_{3}$ and (b) Ni$_{3}$ molecules. The wavefunctions of the down levels at the Fermi energy are displayed.}
\end{figure*}

Figure \ref{figure3b} displays the electronic levels for non-spin-orbit calculations of (a) Ni$_{3}$Bz$_{3}$ and (b) Ni$_{3}$ molecules.
For the Ni$_{3}$Bz$_{3}$ molecule, the highest occupied molecular orbital (HOMO) for the minority spin is a doubly degenerated level occupied by an electron. 
The wave functions of these two levels, displayed nearby, show that they are mainly formed by a combination of $d_{xz}$ and $d_{yz}$ atomic orbitals in each Ni atom. The specific orbital composition of those levels is studied by the projected-density-of-states also shown in the Supplemental Material. 
In the case of Ni$_{3}$, there are two minority spin levels degenerated at the Fermi energy as for Ni$_{3}$Bz$_{3}$, however each level is totally filled with an electron. Their wave functions have $d$ orbital contribution with strong \textit{z} component, as for Ni$_{3}$Bz$_{3}$.
These results agree with those of spin-orbit calculations.

\section{Projected density of states}

\begin{figure}[thpb]
      \centering
\includegraphics[clip,width=0.45\textwidth]{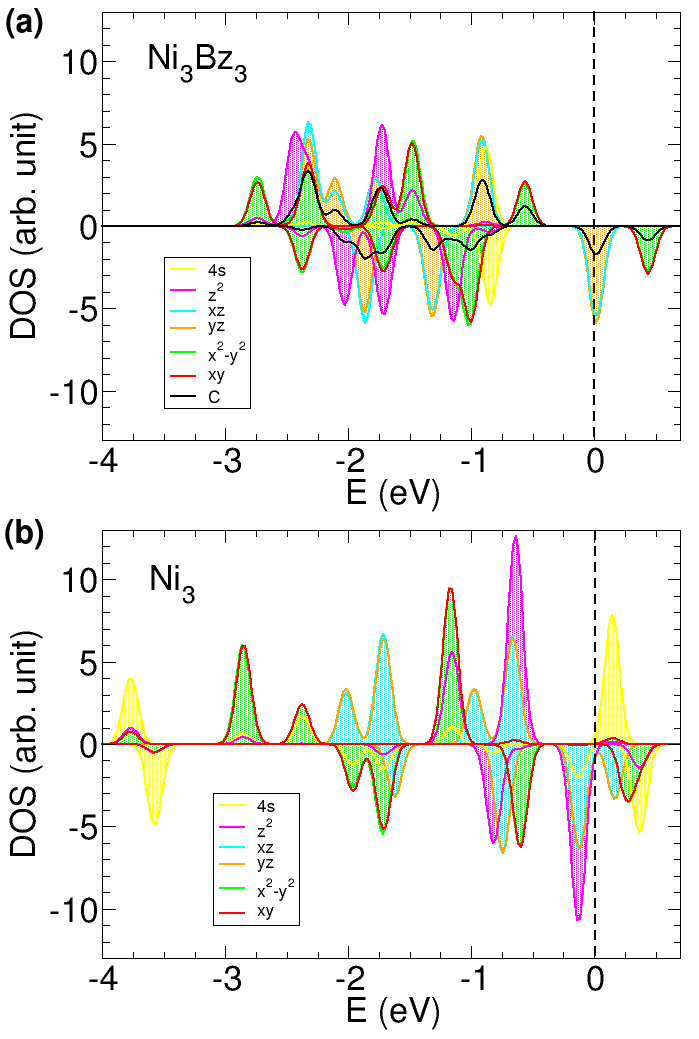}
\caption{\label{figure4} 
Projected density of states for (a) Ni$_{3}$Bz$_{3}$ and (b) Ni$_{3}$.}
\end{figure}

Figure \ref{figure4} shows the projected density of states (PDOS) over C and Ni orbitals for (a) Ni$_{3}$Bz$_{3}$ and (b) Ni$_{3}$ molecules. The coordinate system is the same as used in the main article.
We first comment the PDOS results of Ni$_{3}$Bz$_{3}$ in panel (a). The spin splitting for the d levels near the Fermi energy is estimated about 1 eV.
Carbon contributions expand over almost all the energies, hybridizing with nickel orbitals. The Ni d orbital contributions are brought together into couples, such as 3d$_{xz}$-3d$_{yz}$ and 3d$_{xy}$-3d$_{x^{2}-y^{2}}$, because the hy\-bri\-di\-zed mo\-le\-cu\-lar orbitals have contributions with similar weights at the same energies. The  couple 3d$_{xz}$-3d$_{yz}$ form molecular orbitals in the out-of-plane z-direction, and the second couple 3d$_{xy}$-3d$_{x^{2}-y^{2}}$ forms orbitals in the Ni atom plane. 
Near the Fermi energy, there are two minority spin le\-vels composed of hybridizing 3d$_{xz}$-3d$_{yz}$ Ni orbitals, which are occupied by one electron and are responsible for the magnetic anisotropy of Ni$_{3}$Bz$_{3}$.

Second we discuss the projected density of states for Ni$_{3}$ (see Fig. \ref{figure4}(b)). The scheme of hybridization between \textit{d} orbitals is the same as for Ni$_{3}$Bz$_{3}$. However, the two occupied degenerated levels near the Fermi energy are totally filled, as shown in the level schemes in the main article. The contributions of 4s and 3d$_{z^{2}}$ orbitals show differences between Ni$_{3}$ and Ni$_{3}$Bz$_{3}$ near the Fermi energy. The 4s orbital for Ni$_{3}$ is highly hybridized with the 3d$_{xy}$-3d$_{x^{2}-y^{2}}$ couple and 3d$_{z^{2}}$ orbital contributions.   
The addition of benzene molecules localize a molecular orbital composed of d$_{z^{2}}$ Ni, that for the Ni$_{3}$Bz$_{3}$ molecule is being shifted towards deeper energies together with the 4s contribution. It is noteworthy that for Ni$_{3}$Bz$_{3}$ a level composed of 4s Ni appears about 1 eV below the Fermi energy.


\end{document}